\documentclass{article}
\pdfoutput=1
\usepackage{graphicx}
\usepackage{amsmath}
\usepackage{amssymb}
\usepackage{lscape}

\def\be{\begin{eqnarray}}
\def\ee{\end{eqnarray}}
\def\nn{\nonumber}

\def\tr{{\rm tr}\,}
\def\Tr{{\rm Tr}\,}

\def\l[{\phantom.[}

\def\bS{\bar S}
\def\bT{\bar T}

\def\vac{\emptyset}

\hoffset= - 1.0in         

\begin{document}

\title{{\bf {Tabulating  knot polynomials
for arborescent knots
}\vspace{.2cm}}
\author{{\bf A. Mironov$^{a,b,c,d}$}, \ {\bf A. Morozov$^{b,c,d}$}, \ {\bf An. Morozov$^{b,c,d,e}$}, \\   {\bf P. Ramadevi$^{f}$}, \ {\bf Vivek  Kumar Singh$^{f}$},\
{\bf A. Sleptsov$^{b,c,d,e}$}
}
\date{ }
}

\maketitle

\vspace{-5.5cm}

\begin{center}
\hfill FIAN/TD-01/16\\
\hfill IITP/TH-01/16\\
\hfill ITEP/TH-02/16\\
\end{center}
\vspace{4.2cm}
\begin{center}
$^a$ {\small {\it Lebedev Physics Institute, Moscow 119991, Russia}}\\
$^b$ {\small {\it ITEP, Moscow 117218, Russia}}\\
$^c$ {\small {\it National Research Nuclear University MEPhI, Moscow 115409, Russia }}\\
$^d$ {\small {\it Institute for Information Transmission Problems, Moscow 127994, Russia}}\\
$^e$ {\small {\it Laboratory of Quantum Topology, Chelyabinsk State University, Chelyabinsk 454001, Russia }}\\
$^f$ {\small {\it Department of Physics, Indian Institute of Technology Bombay, Mumbai 400076, India}}
\end{center}
\vspace{1cm}
\begin{abstract}
Arborescent knots are the ones which can be represented in terms of double fat graphs
or equivalently as tree Feynman diagrams.
This is the class of knots for which the present knowledge is enough for
lifting topological description  to the level of effective analytical formulas.
The paper describes the origin and structure of the new tables of colored knot polynomials,
which will be posted at the dedicated site \cite{knotebook}.
Even if formal expressions are known in terms of
modular transformation matrices,
the computation in finite time requires additional ideas.
We use the "family" approach,
suggested in \cite{MMfam}, and apply it to arborescent knots in Rolfsen table
by developing a Feynman diagram technique, associated with an auxiliary matrix model
field theory.
Gauge invariance in this theory helps to provide meaning
to Racah matrices in the case of non-trivial multiplicities
and explains the need for peculiar sign prescriptions in the calculation
of $[21]$-colored HOMFLY polynomials.\end{abstract}
\vspace{1cm}
\section{Introduction}
Chern-Simons field theory \cite{CS} gives a natural setting for
the description of knots in three dimensional space \cite{WitCS}.
Observables in this theory are  expectation values of Wilson loop operators
along knots, which provide knot invariants.
The challenge is to compute the polynomial form \cite{knotpols,Con}
of these invariants, carrying arbitrary representation $R$ of the gauge group
(these are usually called colored HOMFLY-PT   and Kauffman polynomials
for the gauge groups $SU(N)$ and $SO(N)/Sp(N)$ respectively).
Though the methodology is straightforward  in writing the formal
expressions for colored HOMFLY-PT, in terms of modular transformation matrices
$S$ and $T$ and their conjugates (at least for arborescent knots \cite{newinds,gmmms,mmms,mmmrs}), the
calculations are tedious. Moreover, explicit form for Racah matrices $S$ is presently known
only for all symmetric and antisymmetric representations \cite{nrz,mmms} and
for just one  mixed ($[21]$-colored) representation \cite{GJ}.

Colored HOMFLY-PT are believed to be exhaustive knot invariants for the space $S^3$,
while for non-simply-connected spaces one expects much more:
say, HOMFLY-PT  for virtual knots \cite{virtHOMFLY}.
In fact, other polynomials in $S^3$ including Kauffman are believed to
be deducible from colored HOMFLY.
More interesting, for  adjoint representations and their descendants
(the so-called $E_8$-sector of representation theory),
there is an evidence for Vogel's "universality" \cite{Univ},
when many quantities for different groups are described by the same formulas,
depending symmetrically on three parameters, and particular groups appear at their
particular values.
Surprisingly or not, {\it unoriented} knot invariants are exactly the quantities of this type
(while this is probably not quite so for  generic representation).
We observe that the relation of sophisticated superpolynomials (including Khovanov-Rozansky and Floer)  to colored HOMFLY-PT is more obscure: in certain cases, like separation
of mutants, uncolored Floer is already effective while the colored HOMFLY-PT
involving a non-rectangular representation like $[21]$ or $[42]$ must be computed.
Hence we consider the
evaluation of colored HOMFLY-PT  as a very important problem for modern science.
We will show for a class of arborescent knots,  we essentially need the explicit
expression  for the  $S$-matrix elements, specially for  non-rectangular $SU(N)$ representation admitting multiplicities,
to write the polynomial form.

Unfortunately, evaluation of the  $S$ matrix for non-rectangular representation is a very hard problem. One of the
underlying reason is that the Racah matrices for such representation becomes basis-dependent, due to multiplicities, leading
to difficulties in determining them.
It is one of the purposes of this paper to provide an invariant description of the problem:
in terms of auxiliary field theory (matrix model), where ambiguity in the definition of $S$
turns into the ordinary gauge invariance.
Specifics of the problem is that in order to have a {\it gauge invariant}
representation of  {\it knot invariants} one need somewhat non-trivial
 {\it double-fat}  vertices in the Feynman diagrams in the auxiliary field theory.
Alternatively one can work in special gauges which is what we actually follow in our practical calculations. We believe that an understanding of this non-trivial relation between knot and representation theory will provide new insights in both the fields.

Presently,  the study of knot polynomials is actually a field
of "experimental mathematical physics"  and at this stage
having more explicit examples (data), we can empirically
discover  more properties.
Recent advances in knot polynomial calculus in
\cite{inds,MMMkn2,RTmod,newinds,gukovs,mmmrs}
already led to discovery of various non-trivial recursions
\cite{Gar,IMMMfe, MMeqs,GarnonJones}
and factorizations \cite{Kon}.
Certain steps are made towards matrix model reformulation
\cite{BEMS,AMMM}.  Definitely our
attempt at the tabulation of colored knot polynomials data will be
useful to verify known properties and also unearth new properties.

The wonderful knot-database \cite{katlas} and its
descendants \cite{amazing,virknotsite} do not include the most interesting
{\it colored} HOMFLY-PT polynomials. Hence there is a need for the colored HOMFLY-PT  data
leading us to take up the dedicated project \cite{knotebook}.
Once we obtain enormous data in that website, we could tame it in an appropriate way and
include them  in conventional databases.
However, we are very far away from the goal. In fact, there are two comparably
big problems: to calculate knot invariants and to present the results.

Concerning evaluation of colored HOMFLY-PT,
even the calculus for symmetric and antisymmetric  representations
(Young diagrams with a single row or column)
is rather recent \cite{IMMMfe,newinds}, and
the only rich enough example beyond them is the
non-rectangular representation ($[21]$) \cite{AnoCabling,Ano21,GJ,mmmrs,nrs,MMMS21},
where both cabling and $S$-matrix methods can be used.
Some results, but for limited types of non-torus knots,
are also available for rectangular Young diagrams \cite{rect}.
Enlarging the colored HOMFLY-PT for other representations is still an open problem.
A dream could be  to get a formula at least for torus knots, as general as the Rosso-Jones formula
\cite{RJ,China,BEMS,DMMSS},
which describes in a similar way HOMFLY-PT in all representations.

It is important to point out that even the known results on HOMFLY-PT are difficult to use.
As already mentioned, naively these are lengthy formulas,
looking without any pattern. It appears that they should possess much more elegant
reformulations, e.g. via differential   \cite{DGR,evo,arth}
or special-polynomial/Hurwitz  \cite{MMSle,Hurtau} expansions,
as solutions to AMM/EO-like \cite{AMMCEO} recursions
\cite{Gar,Dijk,MMeqs}
and thus represented by matrix models \cite{BEMS,AMMM}.
These  studies are only at the very beginning.
In fact, today the lengthy and almost unstructured answers for knot polynomials
are mainly used in computer programs with the main
aim of  searching  the hidden structures.
This indicates that we need the data in the form convenient for
enumerative analysis.

There are two ways to proceed to list the colored HOMFLY-PT polynomials.
One option is to tabulate the knots and provide formulas for each item
in the list.
This is an attempt, made in \cite{katlas},
based on the Rolfsen table, where knots are ordered according to their
intersection numbers.
This is somewhat tedious both to calculate and to use,
because the computational formula is
written separately for each knot.

Another option, suggested in \cite{MMfam} and which we are going to follow,
is to use the internal structure of knot polynomials themselves and
in every calculational approach tabulate
similar {\it polynomials} (of course, similarity depends in the approach we choose).
Then we  identify the knots described by these formulas
(this stage is simple: one can identify most small knots by their
fundamental HOMFLY and symmetric Jones, which are already tabulated in \cite{amazing}).
In practice, the simplest approach of this kind is to take families
of elements of the braid group, promote them to families of knot polynomials
and then identify knots from the Rolfsen table, that belong to particular
family.  Only hope in this procedure is that, sooner or later,  particular knot will fit into one
or a few  of the families.
If it appears in different families, coincidence of knot polynomials will ensure
evidence for their topological invariance.
This is the approach which we follow in calculations for \cite{knotebook}.

There are two practical ways to make a knot diagram from a braid:
either by taking a trace as in \cite{MMMkn2}, or by taking a matrix element
as in \cite{inds,newinds,mmmrs}.
A typical example of the first kind is a  family of torus knots,
that of the second kind are the  arborescent knots.
A mixed approach of \cite{MMfam},
provides the most general tamed  family of today
that of the "fingered three-strand braids" (F3S family).
Since the main goal of the present paper is to describe our tabulating approach,
we limit consideration to the set of arborescent knots.
These knots are relatively well understood and classified at topological level.
Our goal is to lift this description to analytical level,
where knots are associated with field theory correlators.
In fact, arborescent knots are best prepared for this task, because
the neatest way to describe this set is in terms of peculiar Feynman diagrams (FD),
which we introduce in sec.\ref{FD}.
This language provides a simple way to suggest {\it families},
embracing all knots with a given number of intersections which we explain
in sec.\ref{FDfams}.
In the final  section \ref{beyond} we briefly mention a straightforward
generalization to some knots beyond the arborescent family.
\section{Arborescent (double-fat) knots as Feynman diagrams
\label{FD}}
\subsection{Double-fat knot diagrams
\label{knotdiags} }
Consideration of knots makes sense from  three different approaches  as concisely put forth in the table below.
 In particular, closer focus on these analytical/algebraic description highlights additional structures like those related to orientation.

\centerline{
\begin{tabular}{|c|c|c|}
\hline
& non-oriented & oriented \\
\hline
topological &sec.\ref{knotdiags}&\\
\hline
tensorial &sec.\ref{tensor}& \\
\hline
representational &sec.\ref{nonorpols}& sec.\ref{orpols}\\
\hline
\end{tabular}
}

\vskip.5cm
In \cite{mmmrs} we described the knot polynomial calculus for knot diagrams
of a very special kind, which are in fact  often used to {\it define}
the family of arborescent knots.
They are made from 4-strand braids called "propagators" with the strands grouped pairwise:

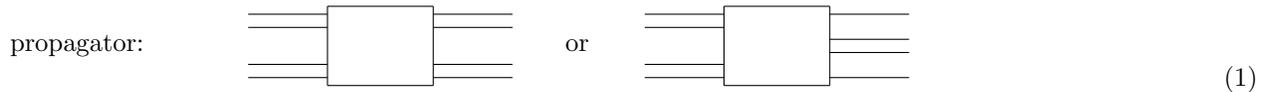
\be
\begin{picture}(300,20)(-30,-15)
\put(-10,-12){\line(1,0){30}}
\put(-10,-7){\line(1,0){30}}
\put(-10,7){\line(1,0){30}}
\put(-10,12){\line(1,0){30}}
\put(60,12){\line(1,0){30}}
\put(60,7){\line(1,0){30}}
\put(60,-7){\line(1,0){30}}
\put(60,-12){\line(1,0){30}}
\put(20,15){\line(1,0){40}}
\put(20,-15){\line(1,0){40}}
\put(20,15){\line(0,-1){30}}
\put(60,15){\line(0,-1){30}}

\put(-100,-2){\mbox{propagator: }}

\put(150,0)
{
\put(-10,-12){\line(1,0){30}}
\put(-10,-7){\line(1,0){30}}
\put(-10,7){\line(1,0){30}}
\put(-10,12){\line(1,0){30}}
\put(60,12){\line(1,0){30}}
\put(60,2.5){\line(1,0){30}}
\put(60,-2.5){\line(1,0){30}}
\put(60,-12){\line(1,0){30}}
\put(20,15){\line(1,0){40}}
\put(20,-15){\line(1,0){40}}
\put(20,15){\line(0,-1){30}}
\put(60,15){\line(0,-1){30}}

\put(-40,-2){\mbox{or }}
}
\end{picture}
\ee
where the first and the fourth strands in the second picture
are considered "close" (as if they were drawn on a cylinder).

Propagators can be attached to planar "vertices" of arbitrary valence
by connecting these pairs of strands. For example, see the following
picture denoting a vertex of valence $4$:

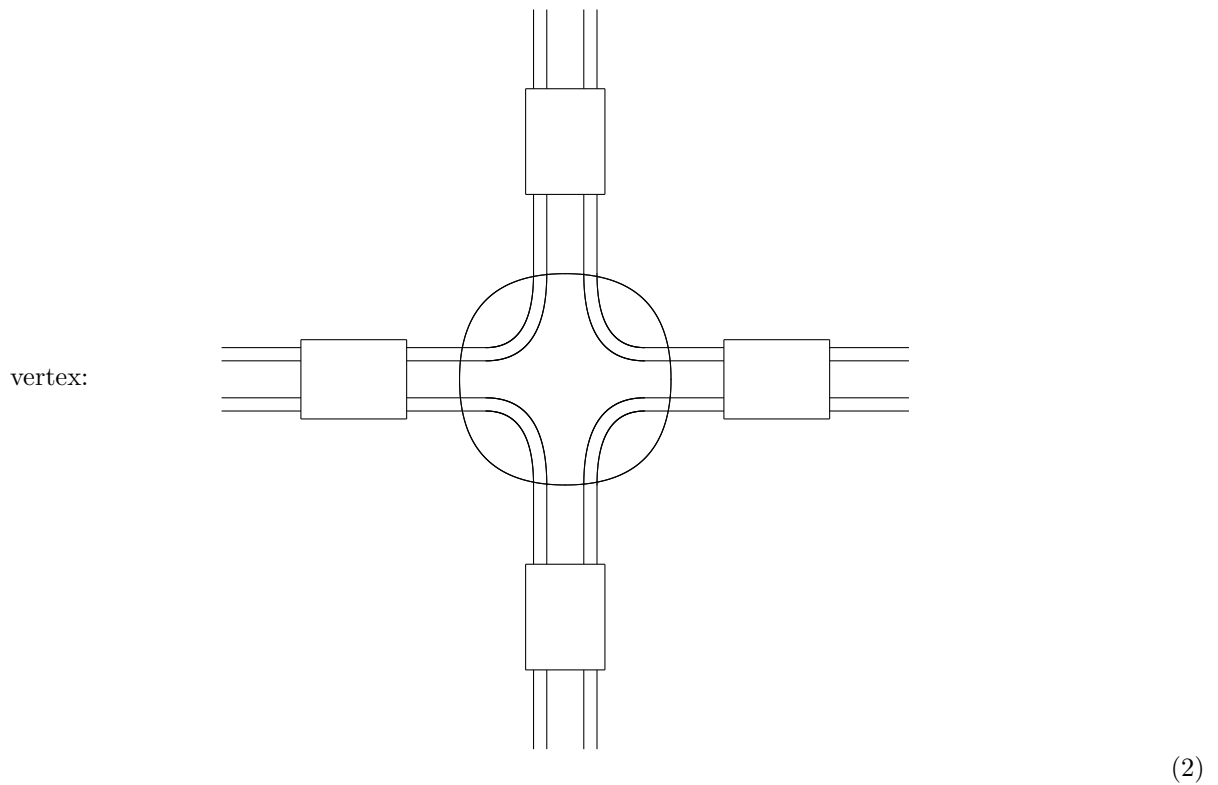
\be
\begin{picture}(300,290)(-200,-150)
\put(0,0)
{
\put(-10,-12){\line(1,0){30}}
\put(-10,-7){\line(1,0){30}}
\put(-10,7){\line(1,0){30}}
\put(-10,12){\line(1,0){30}}
\put(60,12){\line(1,0){30}}
\put(60,7){\line(1,0){30}}
\put(60,-7){\line(1,0){30}}
\put(60,-12){\line(1,0){30}}
\put(20,15){\line(1,0){40}}
\put(20,-15){\line(1,0){40}}
\put(20,15){\line(0,-1){30}}
\put(60,15){\line(0,-1){30}}
}

\put(-160,0)
{
\put(-10,-12){\line(1,0){30}}
\put(-10,-7){\line(1,0){30}}
\put(-10,7){\line(1,0){30}}
\put(-10,12){\line(1,0){30}}
\put(60,12){\line(1,0){30}}
\put(60,7){\line(1,0){30}}
\put(60,-7){\line(1,0){30}}
\put(60,-12){\line(1,0){30}}
\put(20,15){\line(1,0){40}}
\put(20,-15){\line(1,0){40}}
\put(20,15){\line(0,-1){30}}
\put(60,15){\line(0,-1){30}}
}

\put(-40,50)
{
\put(-12,-10){\line(0,1){30}}
\put(-7,-10){\line(0,1){30}}
\put(7,-10){\line(0,1){30}}
\put(12,-10){\line(0,1){30}}
\put(12,60){\line(0,1){30}}
\put(7,60){\line(0,1){30}}
\put(-7,60){\line(0,1){30}}
\put(-12,60){\line(0,1){30}}
\put(15,20){\line(0,1){40}}
\put(-15,20){\line(0,1){40}}
\put(15,20){\line(-1,0){30}}
\put(15,60){\line(-1,0){30}}
}

\qbezier(-10,7)(-33,7)(-33,40)
\qbezier(-10,12)(-28,12)(-28,40)

\qbezier(-70,12)(-52,12)(-52,40)
\qbezier(-70,7)(-47,7)(-47,40)

\qbezier(-10,-7)(-33,-7)(-33,-40)
\qbezier(-10,-12)(-28,-12)(-28,-40)

\qbezier(-70,-12)(-52,-12)(-52,-40)
\qbezier(-70,-7)(-47,-7)(-47,-40)

\put(-40,-130)
{
\put(-12,-10){\line(0,1){30}}
\put(-7,-10){\line(0,1){30}}
\put(7,-10){\line(0,1){30}}
\put(12,-10){\line(0,1){30}}
\put(12,60){\line(0,1){30}}
\put(7,60){\line(0,1){30}}
\put(-7,60){\line(0,1){30}}
\put(-12,60){\line(0,1){30}}
\put(15,20){\line(0,1){40}}
\put(-15,20){\line(0,1){40}}
\put(15,20){\line(-1,0){30}}
\put(15,60){\line(-1,0){30}}
}

\put(-80,0){
\qbezier(0,0)(0,40)(40,40)
\qbezier(0,0)(0,-40)(40,-40)
\qbezier(80,0)(80,40)(40,40)
\qbezier(80,0)(80,-40)(40,-40)
}

\put(-250,-2){\mbox{vertex: }}

\end{picture}
\label{vert}
\ee

\noindent
and can be ended with "cups" (then the propagator becomes a "finger"):

\be
\begin{picture}(100,35)(10,-15)
\put(-170,-2){\mbox{finger  =  propagator with a cup:  }}
\put(0,0)
{
\put(-10,-12){\line(1,0){30}}
\put(-10,-7){\line(1,0){30}}
\put(-10,7){\line(1,0){30}}
\put(-10,12){\line(1,0){30}}
\put(60,12){\line(1,0){30}}
\put(60,7){\line(1,0){30}}
\put(60,-7){\line(1,0){30}}
\put(60,-12){\line(1,0){30}}
\put(20,15){\line(1,0){40}}
\put(20,-15){\line(1,0){40}}
\put(20,15){\line(0,-1){30}}
\put(60,15){\line(0,-1){30}}
}


\qbezier(90,12)(100,12)(100,9.5)
\qbezier(90,7)(100,7)(100,9.5)
\qbezier(90,-7)(100,-7)(100,-9.5)
\qbezier(90,-12)(100,-12)(100,-9.5)

\put(100,0){\circle{40}}

\put(128,-2){\mbox{or}}

\put(160,0)
{
\put(-10,-12){\line(1,0){30}}
\put(-10,-7){\line(1,0){30}}
\put(-10,7){\line(1,0){30}}
\put(-10,12){\line(1,0){30}}
\put(60,12){\line(1,0){30}}
\put(60,2.5){\line(1,0){30}}
\put(60,-2.5){\line(1,0){30}}
\put(60,-12){\line(1,0){30}}
\put(20,15){\line(1,0){40}}
\put(20,-15){\line(1,0){40}}
\put(20,15){\line(0,-1){30}}
\put(60,15){\line(0,-1){30}}



\qbezier(90,12)(105,12)(105,0)
\qbezier(90,2.5)(100,2.5)(100,0)
\qbezier(90,-2.5)(100,-2.5)(100,0)
\qbezier(90,-12)(105,-12)(105,0)


\put(100,0){\circle{40}}
}

\end{picture}
\ee

\noindent
Topologically the propagator can be substituted by a strip. The set of joined propagators
looks just as a fat graph which denotes the Feynman diagram in the auxiliary matrix model.
If we allow only {\it tree} diagrams, then what we obtain in this way are exactly
the knot diagrams of {\it arborescent knots} \cite{Con,Caudron,BS}.
Allowing loops like in \cite{MMfam}, one obtains some non-arborescent knots as well. In this paper, we will briefly highlight the generalization to non-arborescent knots.

\subsection{Tensorial calculus for double-fat (arborescent) knots
\label{tensor}}

As the next step towards constructing knot polynomials, we associate with the propagators the rank-(2,2) tensors
${\Pi}_{AB|CD}$, where each index corresponds to a pair of strands:

\be
\begin{picture}(100,20)(-30,-15)
\put(-10,-12){\line(1,0){30}}
\put(-10,-7){\line(1,0){30}}
\put(-10,7){\line(1,0){30}}
\put(-10,12){\line(1,0){30}}
\put(60,12){\line(1,0){30}}
\put(60,7){\line(1,0){30}}
\put(60,-7){\line(1,0){30}}
\put(60,-12){\line(1,0){30}}
\put(20,15){\line(1,0){40}}
\put(20,-15){\line(1,0){40}}
\put(20,15){\line(0,-1){30}}
\put(60,15){\line(0,-1){30}}

\put(-20,15){\mbox{$A$}}
\put(-20,-20){\mbox{$B$}}
\put(92,15){\mbox{$C$}}
\put(92,-20){\mbox{$D$}}

\put(-100,-2){\mbox{$\Pi_{AB|CD}$:}}

\end{picture}
\ee

\noindent
The indices in vertices and cups are contracted in the obvious way,
and this contraction provides the knot polynomial
for the knot associated with the given diagram.

Tensor  $\Pi_{AB|CD}$ depends on the 4-strand braid inside the box:
it is a contraction of rank-(2,2) tensors, standing at each crossing of adjacent strands.
In fact, there are three different kinds of crossings: $12$, $23$ and $34$,
and the typical formula looks like

\be
\begin{picture}(300,42)(-210,-22)
\put(-10,-12){\line(1,0){90}}
\put(-10,-7){\line(1,0){40}}
\put(-10,7){\line(1,0){40}}
\put(-10,12){\line(1,0){40}}

\qbezier(30,12)(50,12)(70,0)
\qbezier(70,0)(90,-12)(110,-12)

\qbezier(30,-7)(50,-7)(70,0)
\qbezier(70,0)(90,7)(110,7)

\qbezier(30,7)(35,7)(45,9.5)
\qbezier(45,9.5)(55,12)(60,12)

\qbezier(110,-7)(105,-7)(95,-9.5)
\qbezier(95,-9.5)(85,-12)(80,-12)

\put(60,12){\line(1,0){90}}
\put(110,7){\line(1,0){40}}
\put(110,-7){\line(1,0){40}}
\put(110,-12){\line(1,0){40}}

\put(20,20){\line(1,0){100}}
\put(20,-20){\line(1,0){100}}
\put(20,20){\line(0,-1){40}}
\put(120,20){\line(0,-1){40}}

\put(-20,15){\mbox{$A$}}
\put(-20,-20){\mbox{$B$}}
\put(150,15){\mbox{$C$}}
\put(150,-20){\mbox{$D$}}

\put(55,25){\mbox{$A'$}}
\put(80,-30){\mbox{$D'$}}

\put(-280,-2){\mbox{$\Pi_{ABCD} = \sum_{A',D'} K^\pm_{AA'} {\cal K}^\pm_{A'B|CD'} \tilde K^\pm_{D'D}$}}

\end{picture}
\ee
where tilde refers to the second pair of lines.
Propagator of the second type is a similar braid combined with the additional "regrouping operation"

\be
\begin{picture}(300,42)(-130,-22)
\put(-10,-12.5){\line(1,0){130}}
\put(-10,-7.5){\line(1,0){40}}
\put(-10,7.5){\line(1,0){40}}
\put(-10,12.5){\line(1,0){130}}

\qbezier(30,7.5)(40,7.5)(50,5)
\qbezier(50,5)(60,2.5)(70,2.5)

\qbezier(30,-7.5)(40,-7.5)(50,-5)
\qbezier(50,-5)(60,-2.5)(70,-2.5)

\put(70,2.5){\line(1,0){40}}
\put(70,-2.5){\line(1,0){40}}

\put(20,20){\line(1,0){70}}
\put(20,-20){\line(1,0){70}}
\put(20,20){\line(0,-1){40}}
\put(90,20){\line(0,-1){40}}

\put(-20,16){\mbox{$A$}}
\put(-20,-21){\mbox{$B$}}
\put(115,-2){\mbox{$C$}}
\put(120,16){\mbox{$D$}}
\put(120,-21){\mbox{$D$}}

\put(-100,-2){\mbox{$S_{AB|CD}:$}}
\end{picture}
\label{regroup}
\ee
where $D$ refers to the {\it pair} of strands.

It is now natural to impose a requirement that the rank-(2,2) tensor ${\cal K}$ is actually made from
the rank-(1,1) tensor $K$ by conjugation with the same "orthogonal" rank-(2,2)
regrouping tensor ${\cal S}$:
\be
{\cal K}_{AB|CD}^{\pm} =
\sum_{E,F,G} {\cal S}_{AB|EG}K^{\pm}_{EF}{\cal S}_{CD|FG} \nn \\
\sum_{E,G} {\cal S}_{AB|EG} {\cal S}_{CD|EG} = \delta_{AC}\delta_{BD}
\ee
Note that the first formula makes a difference between the third and fourth indices of ${\cal S}$.
Also note that ${\cal S}$ does not commute with the $K$-tensors. Therefore it now matters
 where the regrouping operation is placed in the definition
of the propagator: our convention is to put it at the right end.

A motivation for the above definitions comes from unification \cite{mmmrs} of the modern
(Tanaka-Krein) version \cite{MMMkn2,RTmod} of the Reshetikhin-Turaev (RT) formalism \cite{RT}
with the conformal block calculus of \cite{inds,newinds,gmmms,mmms}.
The former provides the idea of tensor calculus, while the latter is
a concrete suggestion for the definition of caps and fingers.
Originally, the rank-2 tensors $K$ are quantum ${\cal R}$-matrices lifted to the
space of intertwining operators, while ${\cal S}$ are quantum Racah matrices
($6j$-symbols) acting at the same space.
As further noted in \cite{MMfam}, after rank-2 tensors are defined for {\it fingers},
they can be used in the role of $K$ instead of original ${\cal R}$-matrices;
then, different fingers can be inserted into different crossings of the braid.
For {\it fingers}, ${\cal R}$-matrices $K$ is further reduced to
\be
K_{AC}^\pm = T_A^{\pm 1}\delta_{AC},
\label{2red1}
\ee

\subsection{Knot polynomials for non-oriented double-fat (arborescent) knots
\label{nonorpols}}

If one wants to define a knot polynomial in a self-conjugate representation $R$ of Lie algebra ${\cal G}$,
the role of indices $A$ is played by irreducible representations in the tensor square
$R^{\otimes 2}$, and the eigenvalue $K_{AC}$ in (\ref{2red1}) is
\be
T_A = \pm q^{\varkappa_A}
\label{evR}
\ee
where $\varkappa_A$ is the eigenvalue of the second Casimir operator
(i.e. of the cut-and-join operator $\hat W_2$ of \cite{MMN1})
and the sign in (\ref{evR}) depends on whether $A$ belongs to the symmetric or antisymmetric squares
(${\cal S}^2R$ or $\Lambda^2R$) respectively.

Actually, the decomposition is
\be
R^{\otimes 2} =
\oplus_{X} \ W_{X}\otimes X
\label{decoR2}
\ee
where $X$ are irreducible representations, and dimensions of the vector spaces $W_X$
are their multiplicities.
The index $A$ consists of two pieces: $A=(X,\alpha)$, where $\alpha$ labels
elements of some basis in $W_X$.

Now comes the crucial simplification in the theory of {\it arborescent} knots:
{\bf fingers are diagonal in $X$},
\be
F_{(X,\alpha)(Y,\beta)} = \delta_{XY} F_X^{\alpha\beta}
\ee
The reason for this is that  attaching a cup to the 4-strand braid
picks up a singlet representation $\emptyset$ out of $R^{\otimes 4}$, and
\be
\emptyset \in X\otimes Y \ \ \ \Longrightarrow \ \ \  Y=X
\ee
Generic 4-strand braid is labeled  by a sequence of integers
$(l_1,m_1,n_1|l_2,m_2,n_2|\ldots |l_k,m_k,n_k)$ and we define associated finger as
\be
F_X^{(l_1,m_1,n_1| \ldots |l_k,m_k,n_k)}  =
\frac{1}{S_{ X\emptyset}} \sum_{X_1,\ldots, Y_k}
T_X^{l_1}\underbrace{S_{XY_1}T_{Y_1}^{m_1}  S^\dagger_{Y_1X_1}}_{{\cal T}^{m_1}_{XX_1}}  \tilde T_{X_1}^{n_1}
\ \  \ldots \ \
T_{X_{k-1}}^{l_k}\underbrace{S_{X_{k-1}Y_k}T_{Y_k}^{m_k}  S^\dagger_{Y_kX_k}}_{{\cal T}^{m_k}_{X_{k-1}X_k}}
  \tilde T_{X_k}^{n_k}\  \P S_{ X_{K}\emptyset}
 \label{finger}
\ee
All $T$ and $S$ here are still matrices in the Greek indices,
with the only exception of
\be
S_{(X,\alpha),(X,\beta)|\emptyset,\emptyset} = \delta_{\alpha\beta} S_{X\emptyset}
\ee
which is proportional to the unit matrix.
Contraction of the Greek indices at the cap is denoted in above formula by $\P$.
Finally, $\tilde T$ stands for $T$ with the transposed Greek indices:
this operation can be non-trivial if one does not care about choosing some special
bases in the intertwiner spaces $W_X$.
In $S^\dagger$ both Latin and pairs of Greek indices are permuted.

In general, the propagator can be in a non-trivial representation $Q\in R^{\otimes 4}$,
then one needs to know many Racah/mixing matrices $S^{(Q)}$ instead of a single $S=S^{(\emptyset)}$
which appeared in (\ref{finger}).
However, if the entire Feynman diagram is {\it tree}, i.e. the knot is {\it arborescent},
then $Q=\emptyset$ in the propagators as well, and one gets essentially the same formula (\ref{finger}):
\be
P_{XY}^{(l_1,m_1,n_1| \ldots |l_k,m_k,n_k)}  =
\frac{1}{S_{X \emptyset }S_{Y\emptyset}} \sum_{X_1,\ldots, Y_k}
T_X^{l_1}\underbrace{S_{XY_1}T_{Y_1}^{m_1}  S^\dagger_{Y_1X_1}}_{{\cal T}^{m_1}_{XX_1}} \tilde T_{X_1}^{n_1}
\ \  \ldots \ \
T_{X_{k-1}}^{l_k}\underbrace{S_{X_{k-1}Y_k}T_{Y_k}^{m_k}S^\dagger_{Y_kX_k }}_{{\cal T}^{m_k}_{X_{k-1}X_k}}
  \tilde T_{X_k}^{n_k}\ S_{X_kY}
 \label{prop}
\ee
We denote it $P$ instead of $\Pi$ in order to emphasize that it can be used only in {\it trees},
and it is a (1,1)-tensor in $X,Y$, while is still a (2,2)-tensor in the Greek indices.
An additional $S$ at the right end of the propagator appears if one wishes a regrouping.

If the normalization factors $S_{X\emptyset}$ are put in denominators,
as we did in (\ref{finger}) and (\ref{prop}), then at the vertices (\ref{vert}) of the {\it tree}
Feynman diagrams we simply convert the Greek (multiplicity) indices and sum over a single representation
index $X$ with the weight $d_X=S_{\emptyset X}^2 = S_{X\emptyset}^2$, which is just the
quantum dimension of representation $X$ independently of valence of the diagram.
If normalization factors are omitted from (\ref{finger}) and (\ref{prop}), then
the vertex of valence $n$ includes a weight $S_{\emptyset X}^{2-n}$ in the sum over $X$.

{\underline{The last ingredient is the common factor $d_R$ or $d_R^2$ in the case of reduced or non-reduced
knot polynomials respectively.} This factor is needed to make the entire expression
a polynomial.

Putting things together, the arborescent knot described by the {\it tree} Feynman fat diagram
(cf. \cite{Caudron})

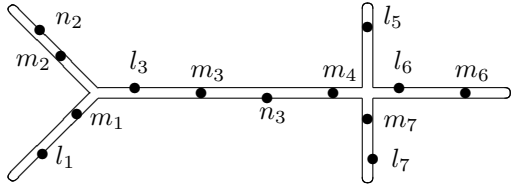
\be
\begin{picture}(300,100)(-150,-50)

\put(0,2){\line(1,0){50}}
\put(0,-2){\line(1,0){50}}
\put(0,2){\line(0,1){30}}
\put(-4,2){\line(0,1){30}}
\put(0,-2){\line(0,-1){30}}
\put(-4,-2){\line(0,-1){30}}
\put(-4,2){\line(-1,0){100}}
\put(-4,-2){\line(-1,0){100}}
\put(-104,2){\line(-1,1){30}}
\put(-104,-2){\line(-1,-1){30}}
\put(-107,0){\line(-1,1){30}}
\put(-107,0){\line(-1,-1){30}}
\qbezier(-4,32)(-4,34)(-2,34)
\qbezier(0,32)(0,34)(-2,34)
\qbezier(-4,-32)(-4,-34)(-2,-34)
\qbezier(0,-32)(0,-34)(-2,-34)
\qbezier(50,2)(52,2)(52,0)
\qbezier(50,-2)(52,-2)(52,0)
\qbezier(-134,32)(-136,34)(-137,33)
\qbezier(-137,30)(-139,32)(-137,33)
\qbezier(-134,-32)(-136,-34)(-137,-33)
\qbezier(-137,-30)(-139,-32)(-137,-33)
\put(-100,0)
{
\put(10,2){\circle*{4}} \put(8,9){\mbox{$l_3$}}
\put(35,0){\circle*{4}} \put(31,6){\mbox{$m_3$}}
\put(60,-2){\circle*{4}} \put(57,-10){\mbox{$n_3$}}
\put(85,0){\circle*{4}} \put(81,6){\mbox{$m_4$}}
\put(110,2){\circle*{4}} \put(108,9){\mbox{$l_6$}}
\put(135,0){\circle*{4}} \put(131,6){\mbox{$m_6$}}
\put(98,25){\circle*{4}} \put(104,26){\mbox{$l_5$}}
\put(100,-25){\circle*{4}} \put(107,-28){\mbox{$l_7$}}
\put(98,-10){\circle*{4}} \put(104,-13){\mbox{$m_7$}}
\put(-18,14){\circle*{4}} \put(-35,10){\mbox{$m_2$}}
\put(-26,24){\circle*{4}} \put(-20,26){\mbox{$n_2$}}
\put(-12,-8){\circle*{4}} \put(-7,-12){\mbox{$m_1$}}
\put(-25,-23){\circle*{4}} \put(-20,-27){\mbox{$l_1$}}
}
\end{picture}
\ee

\noindent
has the colored HOMFLY-PT polynomial

\be
H_R = d_R \sum_{X,Y}d_Xd_Y \sum_{\stackrel{\alpha,\beta,\gamma}{\alpha',\beta',\gamma',\delta'}}
F^{(0,m_1,0|l_1,0,0)}_{X,\beta\gamma} F^{(0,m_2,n_2)}_{X,\alpha\beta}\
P^{(l_3,m_3,n_3|0,m_4,0)}_{X,\gamma\alpha|Y,\alpha'\beta'}\
F^{(l_5,0,0)}_{Y,\beta'\gamma'} F^{(l_6,m_6,0)}_{Y,\gamma'\delta'}
F^{(0,m_7,0|l_7,0,0)}_{Y,\delta'\alpha'}
\ee
In the case of a "pure" propagator, when all $l_3=m_3=n_3=m_4=0$,
it is still non-trivial:
\be
P_{XY}^{0} = \frac{S_{XY}}{S_{\emptyset X}S_{\emptyset Y}}
\label{pureprop}
\ee
 and  it is a  (1,1)-rank tensor in $X,Y$ and (2,2)-rank tensor
in the Greek indices.

\bigskip

In variance with $T$, the Racah matrices $S$ depend not only on $X,Y \in R^{\otimes 2}$, but
also on $R$ itself:
they define the associativity (fusion) map
\be
(R\otimes R)\otimes R \ \longrightarrow X\otimes R  \ \stackrel{S_{XY}}{\longleftrightarrow}\
R\otimes Y \ \longleftarrow \ R\otimes (R\otimes R)
\label{assosc}
\ee
Evaluation of these matrices is the main problem in calculation of colored knot polynomials.
Part of the problem is that they depend on the choice of basis in the intertwining/multiplicity
spaces $W_X$ and, being not quite invariant objects,
do not attract the necessary attention in mathematical literature.
According to the eigenvalue hypothesis of \cite{IMMMev}
(see also some facts {\it pro} and  arguments {\it contra} in \cite{mathev}
and in the last paper of \cite{Univ}),
the Yang-Baxter relations of the braid group allow one to express $S_{AB|CD}$ through
the set of "eigenvalues" $\{T_C\}$, though the explicit expression is rarely known yet.

\subsection{Knot polynomials for oriented double-fat (arborescent) knots
\label{orpols}}
Knot invariants of non-oriented knots are not the most general ones:
they are either the Kauffman polynomials associated with the groups $SO$ and $Sp$,
or the HOMFLY polynomials associated with $Sl$, but only
in the self-conjugate representations.
In fact, all these polynomials seem to be unifiable into a general set of
``universal knot polynomials" \cite{Univ}, where the dependence
on the quantum group parameters is lifted to a symmetric dependence on three continuous parameters.
This family, however, is kind of complementary to the ordinary
colored HOMFLY polynomials in $N$-independent representations of $Sl(N)$
(excluding adjoint of $Sl(N)$ :  $Adj=[21^{N-2}]$ and other self-conjugate representations).
Among other things, this means that the possibility to distinguish between arbitrary prime knots
by {\it universal} colored polynomials is less obvious, even for those who
believe that they are distinguishable by generic colored HOMFLY.
Particularly,   mutant knots are not separated by adjoint polynomials and we believe
that they can be distinguished by other representations.
These are some of the driving reasons for our efforts to calculate
the generic HOMFLY polynomial, which is an invariant of the {\it oriented} knot.

In the case of {\it arborescent} knots, this means that one needs 4-strand braids,
where two strands have an opposite orientation to the other two
(for non-arborescent knots there are loops in Feynman diagrams, and restrictions
on orientation remains only in fingers, see \cite{MMfam,MMMS21}).
From the point of view of representation theory, this means that one now
has $R^{\otimes 2}\otimes \bar R^{\otimes 2}$ instead of $R^{\otimes 4}$,
therefore, there are two types of ${\cal R}$-matrices:
$T$ in the channel $R\otimes R$, which we call "parallel",
and $\bar T$ in the channel $R\otimes\bar R$, which we call "antiparallel",
and, hence, two types of the Racah matrices:
\be
(R\otimes R)\otimes \bar R \ \longrightarrow X\otimes R  \ \stackrel{S_{XY}}{\longleftrightarrow}\
R\otimes Y \ \longleftarrow \ R\otimes (R\otimes \bar R) \nn\\
(R\otimes \bar R)\otimes R \ \longrightarrow X\otimes R  \ \stackrel{\bar S_{XY}}{\longleftrightarrow}\
R\otimes Y \ \longleftarrow \ R\otimes (\bar R\otimes R)
\label{assoscorient}
\ee
We do not need  arbitrary  Racah (or mixing \cite{MMMkn2,RTmod}) matrix from (\ref{assosc}) for the study
of  arborescent knots. However for the study beyond this
arborescent family\cite{MMfam,MMMS21}, general  Racah matrix  plays a big role.
We do not consider it in the present paper and we reserve the same notation $S$
for the first case in (\ref{assoscorient}).
Thus, in our notation, $S$ switches between the parallel and antiparallel sectors,
while $\bar S$ takes the antiparallel sector into the antiparallel one.

Representations $X$ and thus the fingers can now also be parallel and antiparallel,
depending on whether $X \in R^{\otimes 2}$ or $\bar X \in R\otimes \bar R$.
Moreover, the vertices (\ref{vert}) can join only parallel or antiparallel fingers,
but the propagators of the $S$-type can join the parallel vertex to the antiparallel one.
The antiparallel vertices can be connected by the $\bar S$-type propagator.
However, there is no propagator to connect directly two parallel vertices:
the only possibility is just to unify them into a single parallel vertex
of bigger valency; in other words, there is only an ultralocal parallel-parallel propagator.

\subsection{Racah matrices}
Now let us discuss gauge properties of these Racah matrices needed for description of the arborescent knots
(i.e. in {\it tree} Feynman diagrams for {\it double-fat} graphs). As we already explained above, for the arborescent knots
we need Racah matrices of a rather special type,
where the final representation is the singlet $\emptyset$:

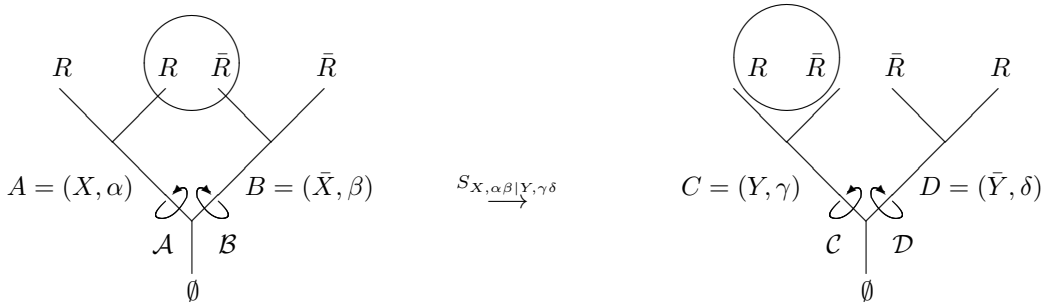
\begin{picture}(200,120)(-100,-15)
\put(0,0){\line(0,1){20}}
\put(0,20){\line(-1,1){50}}
\put(0,20){\line(1,1){50}}
\put(-30,50){\line(1,1){20}}
\put(30,50){\line(-1,1){20}}
\put(-53,75){\mbox{$R$}}
\put(-13,75){\mbox{$ R$}}
\put(7,75){\mbox{$\bar R$}}
\put(47,75){\mbox{$\bar R$}}
\put(-2,-10){\mbox{$\emptyset$}}
\put(-70,30){\mbox{$A=(X,\alpha) $}}
\put(20,30){\mbox{$B = (\bar X,\beta)$}}
\qbezier(-10,27.5)(-17,21)(-11,21)
\qbezier(-11,21)(-6,21.5)(-2,29)
\qbezier(-2,29)(-1,35)(-7,31)
\put(-5.5,32){\vector(-1,-1){2}}
\put(-15,8){\mbox{${\cal A}$}}
\qbezier(10,27.5)(17,21)(11,21)
\qbezier(11,21)(6,21.5)(2,29)
\qbezier(2,29)(1,35)(7,31)
\put(5.5,32){\vector(1,-1){2}}
\put(10,8){\mbox{${\cal B}$}}

\put(0,80){\circle{35}}

\put(100,25){\mbox{$\stackrel{ S_{X,\alpha\beta|Y,\gamma\delta}}{\longrightarrow}$}}

\put(255,0)
{
\put(0,0){\line(0,1){20}}
\put(0,20){\line(-1,1){50}}
\put(0,20){\line(1,1){50}}
\put(-30,50){\line(1,1){20}}
\put(30,50){\line(-1,1){20}}
\put(-45,75){\mbox{$R$}}
\put(-23,75){\mbox{$\bar  R$}}
\put(7,75){\mbox{$\bar R$}}
\put(47,75){\mbox{$R$}}
\put(-2,-10){\mbox{$\emptyset$}}
\put(-70,30){\mbox{$C=(Y,\gamma) $}}
\put(20,30){\mbox{$D = (\bar Y,\delta)$}}
\qbezier(-10,27.5)(-17,21)(-11,21)
\qbezier(-11,21)(-6,21.5)(-2,29)
\qbezier(-2,29)(-1,35)(-7,31)
\put(-5.5,32){\vector(-1,-1){2}}
\put(-15,8){\mbox{${\cal C}$}}
\qbezier(10,27.5)(17,21)(11,21)
\qbezier(11,21)(6,21.5)(2,29)
\qbezier(2,29)(1,35)(7,31)
\put(5.5,32){\vector(1,-1){2}}
\put(10,8){\mbox{${\cal D}$}}

\put(-30,82){\circle{38}}
}
\end{picture}

When there are non-trivial multiplicities in the $X$ and $Y$ channels,
there is an invariance under four independent rotations in the intertwiner spaces
$W_X,W_{\bar X},W_Y,W_{\bar Y}$, acting on indices $\alpha,\beta,\gamma,\delta$ respectively:
\be
S \longrightarrow ({\cal A}\otimes {\cal B}) \ S \ ({\cal C}\otimes {\cal D}) \nn \\
T \longrightarrow  {\cal A} \ T \ {\cal A}^{-1} \nn \\
\tilde T \longrightarrow {\cal B} \ \tilde T \ {\cal B}^{-1}
\ee
or, in more detail,
\be
S_{X,\alpha\beta|Y,\gamma\delta} =
\sum_{\alpha',\beta',\gamma',\delta'}
{\cal A}^X_{\alpha\alpha'}{\cal B}^X_{\beta\beta'}
S_{X,\alpha'\beta'|Y,\gamma'\delta'}{\cal C}^Y_{\gamma\gamma'}{\cal D}^Y_{\delta\delta'}
\ee
where we also explicitly showed that the rotation matrices can depend on representation.
Convolution of $S$'s, $T$'s and $S^\dagger$'s along the braid respects this
"gauge invariance" and provides healthy invariant expressions for the fingers.

It appears that this invariance can be used to diagonalize $S_{\alpha\beta|\gamma\delta}$,
say, in indices $\gamma\delta$.
If this was true, then all fingers could be made commuting and the mutant knots
would remain indistinguishable.
However, $T_{X,\gamma\delta}$ are not quite {\it unit} matrices in $\gamma\delta$,
and this means that the transformation ${\cal C}$ cannot be arbitrary, if $T_X$
is kept diagonal.
This implies that  the gauge freedom is actually smaller (``spontaneously broken") and
one cannot make fingers commuting.

The reason for non-unity of $T_X$ is that the ${\cal R}$-matrix
eigenvalues  for $X_+\in {\cal S}^2R$  and $X_-\in \Lambda^2R$ differ by sign.
In this case, the allowed ${\cal C}$ are arbitrary only in the subspaces $W_{X_+}$
and $W_{X_-}$, and the nondiagonality survives in matrix elements between these
two spaces.

When multiplicity is just two, like it was in the case of $[321]\in [21]^{\otimes 2}$
in ref.\cite{mmmrs}, the only freedom which remains in ${\cal C}$ is the sign:
${\cal C} = \pm I$, i.e. the gauge group reduces from $SO(2)$ to $\mathbb{Z}_2$.
Higher multiplicities, when bigger groups remain unbroken within the symmetric and antisymmetric squares of $R$ respectively
appear starting from $R = [4,2]$ as discussed in Ref.\cite{Morton}.

\subsection{Lagrangian description}

The simplest way to describe and handle the gauge invariance and its consequences
is to reformulate our calculus in terms of some auxiliary Lagrangian.

\paragraph{Fields.} To this end, we introduce the
states/fields:
\be
\begin{picture}(200,30)(0,-12)
\put(0,0){\vector(0,1){20}}
\put(5,0){\vector(0,1){20}}
\put(15,20){\vector(0,-1){20}}
\put(20,20){\vector(0,-1){20}}
\put(8,-13){\mbox{$\sigma$}}
\put(100,0){\vector(0,1){20}}
\put(105,20){\vector(0,-1){20}}
\put(115,0){\vector(0,1){20}}
\put(120,20){\vector(0,-1){20}}
\put(108,-13){\mbox{$\varphi$}}
\put(200,0){\vector(0,1){20}}
\put(205,20){\vector(0,-1){20}}
\put(215,20){\vector(0,-1){20}}
\put(220,0){\vector(0,1){20}}
\put(208,-13){\mbox{$\phi$}}
\end{picture}
\ee
and their conjugates:

\be
\begin{picture}(200,30)(0,-12)
\put(0,0){\vector(0,1){20}}
\put(7.5,0){\vector(0,1){20}}
\put(12.5,20){\vector(0,-1){20}}
\put(20,20){\vector(0,-1){20}}
\put(8,-13){\mbox{$\sigma^*$}}
\put(100,0){\vector(0,1){20}}
\put(107.5,20){\vector(0,-1){20}}
\put(112.5,0){\vector(0,1){20}}
\put(120,20){\vector(0,-1){20}}
\put(108,-13){\mbox{$\varphi^*$}}
\put(200,0){\vector(0,1){20}}
\put(207.5,20){\vector(0,-1){20}}
\put(212.5,20){\vector(0,-1){20}}
\put(220,0){\vector(0,1){20}}
\put(208,-13){\mbox{$\phi^*$}}
\end{picture}
\ee
Each of them carries indices $\sigma_{AB} \longrightarrow \sigma_{X,{\alpha\beta}}$
with the gauge group acting by two orthogonal matrices ${\cal A}$ and ${\cal B}$:
\be
\sigma_{X,\alpha\beta} \ \longrightarrow \
\sum_{\alpha'\beta'}{\cal A}_{\alpha\alpha'}{\cal B}_{\beta\beta'} \sigma_{X,\alpha'\beta'}
\ee

\paragraph{Quadratic terms} in the Lagrangian are:
\begin{itemize}

\item "local" ones
\be
\sigma_XT_X^n\sigma_X = \sigma_{X,\alpha\beta} T^n_{X,\alpha\alpha'} \sigma_{X,\alpha'\beta}
\ \ \ \ \  \varphi_X \bar T^{2n}_X \varphi_X, \ \ \ \
\phi_X \bar  T^{2n}_X \phi_X,
\ \ \ \ \  \varphi_X \bar T^{2n-1}_X \phi_X, \ \ \ \
\phi_X \bar  T^{2n-1}_X \varphi_X
\ee
plus similarly $\sigma_X^* \tilde T_X^n\sigma_X^* = \sigma^*_{X,\alpha\beta}\tilde T^n_{X,\beta\beta'} \sigma^*_{X,\alpha'\beta}
$ etc and plus conjugates,

\item "non-local" ones
\be
\sigma_X^* S_{XY}^\dagger \phi_Y, \ \ \ \ \ \ \phi_X^* S_{XY} \sigma_Y, \ \ \ \ \
\varphi_X^* \bar S_{XY} \varphi_Y
\ee
(note that there are no terms $\phi_X^*\phi_Y$).
\end{itemize}

\paragraph{Vacuum transitions (cups)} are
\be
\bar J\varphi_\emptyset \ \ \ \ {\rm and}   \ \ \ \ J\phi_\emptyset
\ee
note that $\emptyset$ has no multiplicity, hence, no $\alpha\beta$ indices,
thus these vacuum tadpoles do not violate the gauge invariance.

\bigskip

\paragraph{Vertices.} Now one can switch to vertices of our Feynman diagrams.
For concreteness, we will present possible cubic vertex states. It is straightforward to generalize to higher valent vertices.
Topologically allowed are
\be
\Gamma^{(1)}\sim\sigma_X^3, \ \ \ \ \ \ \Gamma^{(2)}\sim\varphi_X^3, \ \ \ \ \ \Gamma^{(3)}\sim\phi_X^2\varphi_X
\label{Lagrverts}
\ee
The problem is, however, to deal with the Greek indices.
A naive anzatz like $\tr \sigma_X^3$ with the trace in Greek indices would be good for
a transformation law $\sigma \longrightarrow {\cal A}\sigma{\cal A}^\dagger$,
but it violates $\sigma \longrightarrow {\cal A}\sigma{\cal B}$ with independent
${\cal A}$ and ${\cal B}$.
This means that at the representational level one cannot get a gauge invariant
description of our knot polynomials.
If one calculates the Feynman diagram for some particular choice of $S$ (in a particular gauge),
the answer differs in other gauges so that there should be some
"handy" compensational rule attached to the answer.

Note that this phenomenon is present even in the absence of multiplicities. Already in the fundamental representation, $R=[1]$ one can use equally well both symmetric and orthogonal Racah matrices, which are related by
\be
S^{symm}_{XY}=\epsilon(X)S^{orth}_{XY}
\ee
giving rise to a factor of $\epsilon(X)^n$ in the $n$-vertex of the Feynman diagram. Here $\epsilon(X)=+1$ or $-1$ for $X\in S^2R$ and $X\in \Lambda^2 R$ respectively.

\subsection{Explanation of sign ambiguity in \cite{mmmrs}}

It turns out that the optimal choice that is applicable also to representations with multiplicities, at least, in the $R=[21]$ case is still to choose the orthogonal matrices. Then, the vertex looks like
\be
\Gamma_{X,\alpha_1,\ldots,\alpha_n}=\prod_i\epsilon(X,\alpha_i)
\ee
and $\epsilon(X,\alpha)=\pm 1$ depending on the representation $(X,\alpha)$ belongs to the symmetric or antisymmetric product of $R\times R$ or $\bar R\times R$ (depending on whether $S$ or $\bar S$ enter the vertex). One can still try to go to symmetric instead of orthogonal Racah matrices in order to remove these $\epsilon$-factors from the vertices.
However, in this case the transition to the symmetric Racah matrices is much less trivial and is given by non-trivial matrices in the multiplicity spaces. This is exactly the phenomenon that we observed in \cite{mmmrs}, expressed there in a ``sign-adjustment" rule.
In that paper, representation $R=[21]$ was considered,
when there is exactly one representation with non-trivial multiplicities. Then, we proposed a gauge choice associated with symmetric Racah matrices, when a non-singlet dependence of the vertex can be reduced to merely a multiplicity dependence, without referring to capital Latin letters. Hence, we considered the interaction in the Lagrangian\footnote{
Invariant formulation should include double-fat vertices of the form
\be
\Gamma^X_{\alpha\alpha',\beta\beta'\gamma\gamma'}\Phi_{X,\alpha'\beta}\Phi_{X,\beta'\gamma}
\Phi_{\gamma',\alpha}
\ee
and only in particular gauges they can be reduced to the ordinary {\it fat-diagram} vertices
\be
\Gamma_{\alpha\beta\gamma}\Phi_{\alpha\beta}\Phi_{\beta\gamma}\Phi_{\gamma\alpha}
\ee
}
\be
\Gamma_{\alpha,\beta,\gamma}\Phi_{\alpha,\beta}\Phi_{\beta,\gamma}\Phi_{X,\gamma\alpha}
\ee
where $\Phi_{\alpha,\beta}$ is a field (any one out of the triple) and we suppressed the $X$-indices in the fields, since the vertex is trivial in these. Now, if one uses $S$ and $\bar S$ from \cite{mmmrs}, 
the vertex can be chosen 1 for the $\sigma^3$ (since, in this case, the corresponding components of fingers with non-unit multiplicities are zeroes), it is cyclically symmetric for $\varphi^3$:
\be
\Gamma_{1,1,1}^{(2)}=\Gamma_{2,2,2}^{(2)}=1;\ \ \ \ \ \ \ \Gamma_{1,1,2}^{(2)}=\Gamma_{2,1,1}^{(2)}=\Gamma_{1,2,1}^{(2)}=\Gamma_{1,2,2}^{(2)}=\Gamma_{2,1,2}^{(2)}=\Gamma_{2,2,1}^{(2)}=-1
\ee
and is more complicated for $\phi^2\varphi$:
\be
\Gamma_{1,1,1}^{(3)}=\Gamma_{2,2,2}^{(3)}=\Gamma_{1,2,1}^{(3)}=\Gamma_{2,1,1}^{(3)}=\Gamma_{2,1,2}^{(3)}=\Gamma_{1,2,2}^{(3)}=1,
\ \ \ \ \ \ \ \ \Gamma_{1,1,2}^{(3)}=\Gamma_{2,2,1}^{(3)}=-1
\ee
where the field $\varphi$ stands at the third places in the vertex: $\Gamma_{\alpha,\beta,\gamma}\phi_{\alpha,\beta}\phi_{\beta,\gamma}\varphi_{\gamma,\alpha}$.

For multiplicities higher than two (i.e. for non-rectangular $R=[42]$
and bigger) the surviving group will not be $\mathbb Z_2$
and the story of Feynman vertices and signs needs to be worked out.

\section{Families of arborescent knots}

\subsection{Abundance of arborescent knots}

As already mentioned, the arborescent knots (which we called "double-fat" in \cite{mmmrs})
are classified \cite{Con,Caudron,BS},
by peculiar  {\it tree} Feynman fat diagrams with two sorts of propagators.
The arborescent set is huge, it includes other popular knot families:
2-strand torus, twist, 2-bridge, pretzel knots. In the
classification of knots with large intersection numbers, we see only a small fraction of {\it all} knots being arborescent.  For example, torus knots with more than two strands
are non-arborescent knots
except the two knots : $8_{19}={\rm Torus}_{[3,4]}$
and $10_{124}={\rm Torus}_{[3,5]}$.

 For small intersection numbers the family of arborescent knots is quite abundant:
the non-arborescent knots in the Rolfsen table (up to 10 crossings) are just
\be
8_{18}, \ \ \underline{9_{34}},\underline{9_{39}},9_{40},9_{41},
\underbrace{9_{47},9_{49}}_{\text{non-alt}}, \ \
\underbrace{10_{100}-10_{123}}_{\text{alternating}}, \underbrace{10_{155}-10_{165}}_{\text{non-alternating}}
\ee
(``alternating" means that there is a knot diagram, where the type of crossing
flips at each step when one walks along the knot, underlined are knots belonging to the
7-parametric family of \cite{MMfam}).

The arborescent knots are distinguished, because in this case one calculates
knot polynomials simply by calculating the Feynman diagrams
inserting appropriate four matrices for $S,\bar S,T,\bar T$,
which depend only on the group and the representation.
In fact, $T$ and $\bar T$ are diagonalized ${\cal R}$-matrices
for parallel and antiparallel lines, while $S$ and $\bar S$ are the
"mixing" (Racah) matrices, converting ${\cal R}$ between the first two strands
in the braid into that between the second and the third strands,
${\cal R}_{23} = S{\cal R}_{12} S^\dagger$.
Again, the choice between $S$ and  $\bar S$
depends on the mutual orientation of strands.
The diagonal $T,\bar T$-matrices are known in full generality (even in the
superpolynomial case \cite{DMMSS}), while the Racah matrices need to be calculated
for any arbitrary representation.
Currently, Racah matrices are known in a universal ($SU(N)$) form only for particular cases of:
the symmetric/antisymmetric representations \cite{gmmms,mmms,nrz} and representation $R=[21]$ as elaborated in Ref.\cite{GJ}.

Each Feynman diagram topology provides a {\it family} of knots in the sense of \cite{MMfam} which is
parameterized by the powers of $T/\bar T$ matrices.
However, these families are not at all independent:
very different Feynman diagrams are equal.
This topological invariance is due to the special (Yang-Baxter) algebraic
properties of the underlying ${\cal R}$-matrices, however,
revealing these equivalencies at the level of Feynman diagrams
made from $S/\bar S$ and $T/\bar T$ is a separate interesting problem.

\subsection{The idea of families}
We will broadly put many knots, where some of them may not be minimal diagrams,
in a family described by some parameters.
By family we mean an {\it evolution} family
of  \cite{evo}, where a ${\cal R}$ matrix, once it appears can be raised to any power,
which is considered as a parameter of the family.
Dependence on these powers is very simple to find in the modern version of the
RT formalism \cite{MMMkn2,RTmod}, which is not the case for dependencies on other possible variations
(like a switch between $S$ and $\bar S$ with accompanying switches $T\leftrightarrow \bar T$).
Looking at evolution families one may have ambitions of different levels:
\begin{itemize}
\item The lowest level is just a technical rule:
once a knot is studied, look at the entire evolution family
and mark everything which fits it. Then, choose the next knot beyond this set.
This provides a systematic approach to quickly exhaust any given set of knots,
and this simple idea turned out enormously effective in calculations of $[21]$-colored
HOMFLY knot polynomials.
\item The intermediate level is an attempt to put all the knots of interest in a single family,
thus getting a description of the entire set by a single formula.
The possibility to proceed this way depends on the meaning of words ``of interest''.
For example, all arborescent knots with less than eight crossings fit
into a rather simple family.
\item The most conceptual level would be getting a new classification of knots,
based on their {\it evolution similarity}.
This means that one can look at the set of knots which are described by families with
certain properties, say, with given topology of Feynman diagram and given finger lengths(length
is dictated by the number of independent powers of ${\cal R}$-matrices involved in the finger).
After that one can search for reasons, why a given knot {\it cannot} fit into a given family
(generalized ``conservation laws"/symmetries).
\end{itemize}

In what follows  we restrict ourselves to the first lowest level.
Instead of a {\it single} formula with just a
few parameters (powers of $T/\bar T$) for all the arborescent knots, say, with no more than 10 crossings,
we have two families that covers practically all of them, still this
is a great simplification both for the knot polynomial calculus
and for presentation of results.
A previous example of this approach is the 7-parametric family of fingered 3-strand knots in \cite{MMfam},
which contains some non-arborescent knots, but at expense of missing quite a lot of
arborescent ones.
Its lifting to a 10-parametric family contains nearly all knots upto 10 crossings,
however, 10 parameters is a little too much.
Though a part of results in \cite{knotebook} is obtained with the use of these 7- and 10-parametric
non-arborescent families,  the majority still comes from the families describing  arborescent ones. These  are somewhat more efficient
for description of  small knots.

\subsection{Building  arborescent families
\label{FDfams}}

The list of minimal tree representations of the arborescent knots up to 11 crossings
can be found in \cite{Caudron}.
Analyzing their structure, one can immediately realize that not too many knots are
described by the pretzel \cite{mmms} and even starfish
(also known as star, or Montesinos) diagrams.
Hence, interesting families contain diagrams with propagators.
It is sufficient  to consider only cubic vertices to describe these knot families
and in what follows we use the two simplest pure propagators
\be
P_{X\bar Y}= \frac{S_{X\bar Y}}{S_{\emptyset X} \bar S_{\emptyset \bar Y}} \ \ \ \
~~{\rm and}\ \ \ \
\bar P_{\bar X\bar Y} = \frac{\bar S_{\bar X\bar Y}}{\bar S_{\emptyset \bar X} \bar S_{\emptyset \bar Y}}
\ee
In our families, we also use the following few short fingers, parallel:
\begin{equation}
\begin{array}{rclc}
F_{ap}^{(n)}&= &\displaystyle{\frac{(\bS \bT^n S)_{\emptyset X}}{S_{\emptyset X}}} &   n \ \text{odd}\cr
\cr
F_{aap}^{(m,n)}&=& \displaystyle{\frac{(\bS\bT^m \bS \bT^n S)_{\emptyset X}}{S_{\emptyset X}}} &
\ \ \ m \ \text{even},\ n \ \text{odd} \cr
\cr
F_{pap}^{(m,n)} &=& \displaystyle{\frac{(ST^mS^\dagger \bT^n S)_{\emptyset X}}{S_{\emptyset X}}} & n \ \text{even}\cr
\ldots
\end{array}
\end{equation}
and antiparallel:
\begin{equation}
\begin{array}{rclc}
F_{pa}^{(n)} &= &\displaystyle{\frac{(ST^nS^\dagger)_{\emptyset \bar X}}{\bS_{\emptyset \bar X}}}  \\
\\
F_{aa}^{(n)} &=& \displaystyle{\frac{(\bS \bT^n \bS)_{\emptyset \bar X}}{\bS_{\emptyset \bar X}}} &   n \ \text{even} \\ 
\\
F_{aaa}^{(m,n)}& =& \displaystyle{\frac{(\bS\bT^m\bS\bT^n\bS)_{\emptyset \bar X}}{\bS_{\emptyset \bar X}}} &
 \ \  m,n \ \text{even} \\
 \\
F_{paa}^{(m,n)} &=& \displaystyle{\frac{(S T^m S^\dagger\bT^n\bS)_{\emptyset \bar X}}{\bS_{\emptyset \bar X}}} &   n \ \text{odd}  \\
\\
F_{apa}^{(m,n)} &=& \displaystyle{\frac{(\bS \bT^m S T^n S^\dagger)_{\emptyset \bar X}}{\bS_{\emptyset \bar X}}} &   m \ \text{odd} \\ \\
F_{apaa}^{(l,m,n)} &=& \displaystyle{\frac{(\bS \bT^l S T^m S^\dagger\bT^n\bS)_{\emptyset \bar X}}{\bS_{\emptyset \bar X}}} &
\ \  l,n \ \text{odd} \\
\\
&&\ldots
\end{array}
\end{equation}
In the cases, when some power parameter can be put to zero, the two adjacent $T$-matrices merge
so that their powers are
added together, while the total number of $T$-insertions drops by two.
In practice, one also has to impose some additional restrictions
on the parity of indices or of their partial sums in order to generate knots, but not links.
Note that using  fat graph diagrams of  knots with shorter fingers drastically simplifies computer
evaluation of the knot polynomials.

Greek indices are suppressed, but they are always present,
at least in some $X$, when $R$ is a non-rectangular representation.
Moreover, as matrices in the Greek indices, the fingers for non-rectangular representations $R$
do not commute, which allows these knot polynomials to distinguish between
mutants.
\subsection{Examples}

We are now ready to provide examples of rather rich families with Feynman diagrams of different topology.
They are rather rich, so it is more practical to list the knots with upto 10 crossings
from the Rolfsen table \cite{katlas}, which do {\it not} get to the family (at least, up to not too large values of parameters in the family, see below).

In fact, it is easy to claim that the knot belongs to the family by checking whether the
fundamental HOMFLY and the $[2]$-colored Jones match with the polynomials listed in \cite{katlas} and \cite{amazing}.
The fundamental HOMFLY alone is not quite enough as there are accidental coincidences, e.g.,

\be
H_{5_1} \stackrel{H_{_\Box}}{\cong} H_{10_{132}} \ \ \ ({\rm i.e.}\
H_{_\Box}^{5_1} = H_{_\Box}^{10_{132}}), \ \ \
8_8 \stackrel{H_{_\Box}}{\cong} 10_{129}, \ \ \ 8_{16}  \stackrel{H_{_\Box}}{\cong} 10_{156},\ \
10_{25} \stackrel{H_{_\Box}}{\cong} 10_{56}, \ \ 10_{40} \stackrel{H_{_\Box}}{\cong} 10_{103}, \nn \\
10_{103} \stackrel{H_{_\Box}}{\cong} 10_{40},\ \
10_{155}\stackrel{H_{_\Box}}{\cong} 11_{n37},\ \ 10_{100} \stackrel{H_{_\Box}}{\cong} ???
\ee
where the last knot, $10_{100}$ which is non-arborescent has the same fundamental HOMFLY as an arborescent knot with intersection not more than 16 and not less than 13 crossings. Note that this $H_{_\Box}$-equivalence preserves the knot property of being (non)alternative.

Certainly, the $[2]$-colored Jones polynomials distinguish the knots in these pairs.
One could suspect that there can be degeneracies in both $H_1$ and $J_2$ with some more
complicated knots, but this is excluded by our restriction on crossing numbers.
Proving that the knot does {\it not} belong to the family is far more complicated:
it can appear at rather high values of evolution parameters, as it actually happens for
many pretzel knots in \cite{mmms}.
Therefore, below we list the arborescent knots which {\it can} be missing in given families.
The concrete values of parameters, providing the knots, which are {\it present} in the families,
are collected at \cite{knotebook}.

The families below are ordered by increasing topology of the diagram,
not by the number of evolution parameters, what looks more interesting conceptually.
However, for the actual computer time the situation is opposite:
it depends more on the number of $S$-matrices,
than on topology: calculations for $[21]$-colored HOMFLY for pretzel knots
(when all fingers are of length one)
are 2-3 orders of magnitude faster than for the families with fingers of length 3.

\begin{itemize}
\item[1)] Feynman diagrams which are {\bf segments} with dots
(one closed finger) describe {\bf rational (2-bridge)} knots \cite{rats}.
They are unambiguously parameterized  by a single
rational number, which should be represented as a continuous fraction
\be
\frac{\alpha}{\beta} =
\frac{1}{n_1+\frac{1}{n_2+\frac{1}{n_3+\frac{1}{n_4+\ldots}}}}
\ee
then the non-oriented knot polynomial is
\be
d_R \Big(ST^{n_1}ST^{-n_2}ST^{n_3}ST^{-n_4}ST^{n_5}\ldots \Big)_{\emptyset\emptyset}
\ee
and bars are uniquely restored in the oriented case.
\item[2)]
{\bf Starfish} Feynman diagrams (one vertex with any number $k$ of parallel or
antiparallel fingers attached) describe {\bf Montesinos} knots \cite{Monte},
parameterized by  sets of rational numbers $\alpha_i/\beta_i, \ \ i=1,\ldots,k$
When they are all integers (all fingers are one-parametric), we get
{\bf pretzel} knots.
Colored HOMFLY in pretzel case were studied in detail in \cite{gmmms,mmms}.
Here we report that the $H_{[21]}$ calculation is finalized for all pretzel knots upto 10 crossings and
the result is posted at \cite{knotebook}.

\item[3)]
The 9-parametric 3-finger {\it starfish} family with more complicated (non-pretzel) fingers
\be
d_R \sum_{\bar X} d_{\bar X} \prod_{i=1}^3 F_{apaa}^{l_i,m_i,n_i}(\bar X)
\ee
could miss the following arborescent knots:

$8_{16},8_{17}, \ \ \ 9_{29},9_{32},9_{33},9_{35},9_{37},9_{38},9_{46},9_{48},$ \ \ \
$10_{67},10_{68},10_{69},10_{74},10_{75}$, \ $10_{79}-10_{99}$ and
$10_{145}-10_{154}$

The same set is missing in more complicated starfish families
and it is very close to the list of knots (underlined) with upto 10 crossings that are claimed not to be Montesinos knots of length at most 3, \cite{p17}.

Also, this family does not contain 11-crossing mutant knots
(despite there are many Montesinos and even pretzel knots among mutants).

\bigskip

\item[4)]
Among the {\bf 4-point tree} Feynman diagrams (with one propagator) we mention the following:

\item 5-parametric $Q_5^{(1)}$:
\be
d_R\sum_{\bar X,\bar Y} d_{\bar X}d_{\bar Y}
\cdot F_{pa}(\bar X)F_{pa}(\bar X) \bar P_{\bar X\bar Y}
F_{paa}(\bar Y)F_{aa}(\bar Y)
\label{Q51}
\ee
It does {\it not} contain:
\be
9_{10}, 9_{17},9_{23},9_{27},9_{28},9_{29},9_{30},9_{31},9_{35},9_{37},9_{38},9_{46},9_{48},\nn\\
 10_{16}, 10_{21},10_{23},10_{27},10_{30},10_{32},10_{33},10_{37},
10_{40},10_{41},10_{42},10_{43},10_{44},10_{45}, 10_{50},10_{51},\nn \\
10_{53},10_{57},10_{58},10_{59},10_{60},10_{64},10_{66},10_{67},10_{68},10_{69}, 10_{71},
10_{73},10_{74},10_{75}, 10_{77},10_{78},10_{81},10_{83},\nn \\
10_{86},10_{88},10_{89},10_{92},10_{95},10_{96},10_{97},10_{98},10_{99},
10_{136},10_{137},10_{138},10_{145},10_{146},10_{147}, 10_{154}
\nn\ee

\noindent
i.e. Thirteen 9-crossing and fifty 10-crossing arborescent knots out of $43$ and $130$
respectively.

\item Another 5-parametric family $Q_5^{(2)}$
\be
d_R\cdot \sum_{\bar X,Y}d_{\bar X}d_Y F_{pa}(\bar X)F_{aa}(\bar X)S_{Y\bar X}F_{ap}(Y)F_{aap}(Y)
\ee
does {\it not} contain:
\be
8_9,8_{10},8_{16},8_{17}, \ \
9_1,  9_{16},9_{17},9_{20},9_{23},9_{26},9_{27},9_{28},
9_{30},9_{31},9_{32},9_{33}, \ \
10_2,10_5, 10_9,10_{12},10_{14}, 10_{17},  \nn \\
10_{21}\!-\!10_{23},10_{26},10_{27},10_{29},10_{32},10_{37},
10_{39}-10_{45}, 10_{48}, 10_{57}\!-\!10_{60},10_{62},10_{64},
10_{66}, 10_{69},
10_{71},10_{72},\nn\\
10_{73},10_{75}\!-\!10_{85},10_{87}\!-\!10_{91},10_{93},10_{94},
10_{96}\!-\!10_{99},   10_{135}\!-\!10_{139},10_{141},10_{143},
10_{148}\!-\!10_{154}
\nn
\ee
\item A 6-parameter family $Q_6^{(1)}$:
\be
d_R\sum_{\bar X, \bar Y}
d_{\bar X} d_{\bar Y} F_{apa}(\bar X) F_{pa}(\bar X) \bar P_{\bar X\bar Y} F_{apa}(\bar Y)F_{pa}(\bar Y)
\label{Q61}\ee
does not contain the following knots:
\be
 9_{28}, 9_{29}, 9_{35}, 9_{37}, 9_{38}, 9_{46}, 9_{48}, \ \ \
  10_{32}, 10_{40}, 10_{42}, 10_{58}, 10_{66}, 10_{67}, 10_{68}, 10_{69}, 10_{71}, 10_{74},
 \nn \\
10_{75},10_{77}, 10_{78}, 10_{82}, 10_{84}, 10_{85}, 10_{87}, 10_{92}, 10_{95}, 10_{96}, 10_{97},
10_{98}, 10_{145}, 10_{146}, 10_{147}
\label{Q61list}\nn
\ee
\item A 7-parameter family $Q_7^{(1)}$
\be
d_R\sum_{\bar X,\bar Y} d_{\bar X} d_{\bar Y}\cdot F_{pa}(\bar X) F_{pa}(\bar X) \bar P_{\bar X\bar Y} F_{paa}(\bar Y)F_{apaa}(\bar Y)
\ee
does not contain the following knots:
\be
9_{29}, 9_{35}, 9_{37}, 9_{38},9_{46}, 9_{48}, \ \ \
10_{58}, 10_{59}, 10_{60}, 10_{67}, 10_{68}, 10_{69}, 10_{74}, 10_{75},
 10_{81}, 10_{83}, 10_{86},
\nn \\
10_{88}, 10_{89}, 10_{92}, 10_{95}, 10_{96}, 10_{97}, 10_{98},
10_{99}, 10_{136}, 10_{137}, 10_{138}, 10_{145}, 10_{146}, 10_{147}, 10_{154}
\nn\ee

Note that $Q_5^{(1)}$  in (\ref{Q51})
is {\it not} a subset of $Q_7^{(1)}$, because the parameters in $F_{apaa}$ are not
allowed to vanish.
\item The best parametric family (for describing upto 10-crossing knots)  in this class
(of 4-point Feynman trees with up to 7 parameters) looks  like family $Q_7^{(2)}$:
\be
d_R\sum_{X,\bar Y} F_{ap}(X)F_{pap}(X)T^n_X \bar P_{X\bar Y} F_{apa}(\bar Y)F_{aa}(\bar Y)
\ee
(notice the additional $T$-insertion).
It does not contain the following knots:
\be
9_{32}, 9_{33}, \ \ \
 10_{45}, 10_{57}, 10_{62}, 10_{64}, 10_{66}, 10_{79}, 10_{80}, 10_{81}, 10_{82},
10_{83}, 10_{84}, 10_{85}, 10_{87}, 10_{88}, 10_{89},
 \nn \\
  10_{90}, 10_{91}, 10_{94},
  10_{98}, 10_{99},10_{139}, 10_{141}, 10_{143}, 10_{148}, 10_{149}, 10_{150}, 10_{151},
10_{152}, 10_{153}, 10_{154}
\label{Q72list}\nn
\ee
\item Going to {\bf 5-point Feynman trees} with two propagators,
we get families, containing the 11-crossing mutant representations from \cite{mmmrs}:
\item A 7-parametric
\be
d_R \sum_{\bar X,Y,\bar Z} d_{\bar X} d_Y d_{\bar Z}\cdot
F_{pa}(\bar X)F_{aa}(\bar X)P_{Y\bar X} F_{ap}(Y)T_Y^{n} P_{\bar ZY} F_{aaa}(\bar Z)F_{pa}(\bar Z)
\label{apTa}
\ee
can miss
\be
 8_{16},
 \ \ \    9_{31},9_{33},
\ \ \    10_{40},10_{42},10_{43}, 10_{45},10_{57},10_{58},10_{60},\nn \\
10_{64},10_{66},10_{71},10_{73},
  10_{79},10_{80},10_{81},10_{83},10_{84},10_{85},10_{88},10_{89},
 10_{91},10_{93}, 10_{99}
\nn\ee
but includes four 11-crossing mutant pairs
\be
11a57/11a231, 11n71/11n75, 11n73/11n74, 11n76/11n78
\nn\ee

\item Amusingly, if  $T_Y^n$ is changed for $\bar T_X^n$ in (\ref{apTa}),
\be
d_R \sum_{\bar X,Y,\bar Z} d_{\bar X} d_Y d_{\bar Z}\cdot
F_{pa}(\bar X)F_{aa}(\bar X)\bar T_X^{n}P_{Y\bar X} F_{ap}(Y) P_{\bar ZY} F_{aaa}(\bar Z)F_{pa}(\bar Z)
\label{aTpa}
\ee
the mutants disappear from such a family,
but instead  at most only 15  arborescent knots from the Rolfsen table
\be
 9_{31}, \ \ 10_{17}, 10_{40}, 10_{42},10_{43},10_{44},10_{45},
10_{60}, 10_{64}, 10_{69}, 10_{75}, 10_{88}, 10_{89}, 10_{98}, 10_{99}
\nn\ee
are missing from it.
It also provides two {\it  new ``false non-arborescent knots"}:
$10_{102} \ \stackrel{H_{_\Box}}\cong \ ???$ and $10_{111} \ \stackrel{H_{_\Box}}\cong \ ???$
that are really some arborescent knots with more than 12 crossings.
\item
A 6-parametric family
\be
d_R \sum_{\bar X,\bar Y,\bar Z} d_{\bar X} d_{\bar Y} d_{\bar Z}\cdot
F_{pa}(\bar X)F_{pa}(\bar X)\bar P_{\bar Y\bar X} F_{aa}(\bar Y)
\bar P_{\bar Z\bar Y} F_{aa}(\bar Z)F_{aa}(\bar Z) \bar T_Z^n
\label{aaa}
\ee
contains 11 pairs of 11-crossing mutants:
\be
&11a19/11a25, \ 11a24/11a26, \ 11a251/11a253, \ 11a252/11a254, \ 11n34/11n42, \nn \\
&11n35/11n43,\ 11n36/11n44, \ 11n39/11n45,\  11n40/11n46,\ 11n41/11n47, \ 11n151/11n152
\ee
Among non-mutants it can miss
\be
9_{10}, 9_{17}, 9_{23}, 9_{26}, 9_{27}, 9_{29}, 9_{31}, 9_{35}, 9_{37}, 9_{38}, 9_{46}, 9_{48} \nn \\
10_{14}, 10_{16}, 10_{21}, 10_{23}, 10_{26}, 10_{27}, 10_{29}, 10_{30}, 10_{32}, 10_{33}, 10_{37},
10_{40}, 10_{41}, 10_{42}, 10_{43}, 10_{44}, 10_{45}, 10_{57}, \nn \\
10_{58}, 10_{59}, 10_{60}, 10_{64}, 10_{66}, 10_{67}, 10_{68}, 10_{69}, 10_{74},
10_{75}, 10_{83}, 10_{86}, 10_{88}, 10_{89}, 10_{92}, 10_{95}, 10_{96}, 10_{97}, \nn \\
10_{98}, 10_{99},  10_{136}, 10_{137}, 10_{138}, 10_{145}, 10_{146}, 10_{147}
\nn\ee
The lacking 11-crossing mutant pairs are among the pretzel mutants:
\be
11a44/11a47, \ 11a57/11a231, \ 11n71/11n75, \ 11n73/11n74, 11n76/11n78
\nn\ee
\item One can add one more propagator and get richer families. For example, represent finger $F_{pa}$ as a propagator like $P_{pa}^{(n)}(X,Y) = \frac{(ST^nS^\dagger)_{\bar Y \bar X}}{\bS_{\bar Y \bar X}}$ and consider the following 8-parametric family $Q_8^{(1)}$:
\be\label{Q81}
d_R\sum_{\bar X,Y,\bar Z,\bar W}d_{\bar X} d_{\bar Y} d_{\bar Z}  d_{\bar W}\cdot P_{pa}(\bar X, \bar Z)F_{paa}(\bar X)P_{\bar XY}F_{ap}(Y)F_{ap}(Y)F_{pa}(\bar Z)\bar P_{\bar Z\bar W}F_{pa}(\bar W)F_{pa}(\bar W)
\ee
It can miss knots only starting from 10 crossings:
\be
10_{43}, 10_{44}, 10_{45}, 10_{66}, 10_{83}, 10_{86}, 10_{88}, 10_{89}, 10_{92}, 10_{95}, 10_{96}, 10_{97} \nn
\ee
\end{itemize}

\noindent
Clearly, the above families  contains all the arborescent knots within the Rolfsen table. In fact just three, say, (\ref{Q61}), (\ref{aTpa}) and (\ref{Q81}) are enough, and they
were actually used in the calculations of colored HOMFLY.
\subsection{Arborescent mutants
\label{mutants}}
Description and separation of mutants is the current important problem in the theory
of knot polynomials.
In knot theory mutation is the transformation of knot diagram,
when one cuts away a box with just four external legs and rotates or reflects it
before gluing back.
As argued in \cite{Mort}, mutants can be separated only by representations $R$
with non-trivial multiplicities in $R^{\otimes 2}$, this means by $R$, which are
non-rectangular Young diagrams, the first of them being $[2,1]$.
Moreover, for the reasons, which are intimately related to our discussion in the
last three subsections of sec.\ref{FD},
some mutants (say, antiparallel pretzels) get separated only by $R$,
where non-trivial multiplicities
appear in symmetric or antisymmetric squares
${\cal S}^2R$ and $\Lambda^2R$ \cite{Morton}. To attempt distinguishing other
mutants (like antiparallel pretzels), we need to go to representation
$R$ where multiplicity is greater than two.
This raises the necessity  to consider $R$ to be at least $[4,2]$.

Evaluation of $[21]$-colored HOMFLY for the simplest 11-crossing mutants
became possible just recently
\cite{mmmrs,nrs}, based on achievement of \cite{GJ}
(though the very fact of separability
was demonstrated by $SU(4)$ calculation of the difference in \cite{Mort} quite some years ago).
But even then all the sixteen 11-crossing mutant pairs were out of reach.
Family approach makes this easy, and  the completion of the table
in \cite{mmmrs} can be found at \cite{knotebook}.

As to $[4,2]$, there is only the basic $SU(3)$-evaluation for the HOMFLY difference
between the simplest pair of pretzel mutants in \cite{Morton}.
Evaluation of the entire $[4,2]$-colored HOMFLY remains a next big challenge
for modern mathematical physics.

Now we are able to present a check of our conjecture \cite{mmmrs} of the universal difference of the $[2,1]$ HOMFLY polynomials for the pairs of mutant knots for all 16 pairs with 11 crossings. The conjecture claims that the difference is
\be
\Delta H_{[2,1]}^{mutant}=A^\gamma\cdot f(A,q)\cdot M(q)
\ee
where $\gamma$ is an integer, $M(q)$ is a function of only $q$, which is a ratio of quantum numbers and
\be
f(A,q):=\{q\}^{11}\cdot  D_{3}^2D_{2} D_0 D_{-2}D_{-3}^2 \nn
\ee
where, as usual, $[...]$ denotes the quantum numbers, $\{x\}\equiv x-1/x$ and $D_k=\{Aq^k\}/\{q\}$.

These differences of $[2,1]$-colored HOMFLY for all sixteen 11-crossing mutant pairs are\footnote{Note that in the concrete checks of the conjecture in \cite{mmmrs} there are misprints: there sometimes mistakenly appears $D_3^3$ instead of correct $D_3^2$.}:
\be\label{mutlist}
1.\qquad H^{11a19}_{[2,1]} - H^{11a25}_{[2,1]} &=& A^{-7}\cdot f(A,q)\cdot\dfrac{[14]}{[2][7]}\cdot\mathfrak{n}   \nn \\
2.\qquad H^{11a24}_{[2,1]} - H^{11a26}_{[2,1]} &=& A^{-1} \cdot f(A,q)\cdot\dfrac{[14]}{[2][7]}\cdot\mathfrak{n}  \nn \\
3.\qquad H^{11a44}_{[2,1]} - H^{11a47}_{[2,1]} &=& A \cdot f(A,q)\cdot\frac{[8]}{[2]}\cdot\mathfrak{n}  \nn \\
4.\qquad H^{11a57}_{[2,1]} - H^{11a231}_{[2,1]} &=& A^{-5}\cdot f(A,q)\cdot\frac{[8]}{[2]}\cdot\mathfrak{n}  \nn \\
5.\qquad H^{11a251}_{[2,1]} - H^{11a253}_{[2,1]} &=& A^{-1} \cdot f(A,q)\cdot\dfrac{[14]}{[2][7]}\cdot\mathfrak{n}  \nn \\
6.\qquad H^{11a252}_{[2,1]} - H^{11a254}_{[2,1]} &=& A^{-5} \cdot f(A,q) \cdot\dfrac{[14]}{[2][7]}\cdot\mathfrak{n} \nn \\
\nn \ee\be
7.\qquad H^{11n34}_{[2,1]} - H^{11n42}_{[2,1]} &=& A^{3} \cdot f(A,q)\cdot\dfrac{[14]}{[2][7]}\cdot\mathfrak{n}   \nn \\
8.\qquad H^{11n35}_{[2,1]} - H^{11n43}_{[2,1]} &=& A^{19} \cdot f(A,q)\cdot\mathfrak{n}  \nn \\
9.\qquad H^{11n36}_{[2,1]} - H^{11n44}_{[2,1]} &=& A^{-9}\cdot f(A,q)\cdot\mathfrak{n}   \nn \\
10.\qquad H^{11n39}_{[2,1]} - H^{11n45}_{[2,1]} &=& A^{-3}\cdot f(A,q)\cdot\dfrac{[14]}{[2][7]}\cdot\mathfrak{n} \nn \\
11.\qquad H^{11n40}_{[2,1]} - H^{11n46}_{[2,1]} &=& A^{13} \cdot f(A,q)\cdot\mathfrak{n}  \nn \\
12.\qquad H^{11n41}_{[2,1]} - H^{11n47}_{[2,1]} &=& A^{-15}\cdot f(A,q)\cdot\mathfrak{n}  \nn \\
13.\qquad H^{11n71}_{[2,1]} - H^{11n75}_{[2,1]} &=& A^{13} \cdot f(A,q)\cdot\dfrac{[7][8]}{[14]}\cdot\mathfrak{n}   \nn \\
14.\qquad H^{11n73}_{[2,1]} - H^{11n74}_{[2,1]} &=& A^{-3}\cdot f(A,q)\cdot\dfrac{[8]}{[2]}\cdot\mathfrak{n}    \nn \\
15.\qquad H^{11n76}_{[2,1]} - H^{11n78}_{[2,1]} &=& A^{-15} \cdot f(A,q)\cdot\dfrac{[7][8]}{[14]}\cdot\mathfrak{n}    \nn \\
16.\qquad H^{11n151}_{[2,1]} - H^{11n152}_{[2,1]}  &=& A^{-9} \cdot f(A,q)\cdot\dfrac{[14]}{[2][7]}\cdot\mathfrak{n}
\ee
where, for the sake of brevity, we introduced a standard factor $\mathfrak{n}:=\dfrac{[3]^2[14]}{[2][7]}$.

\subsection{Beyond arborescent knots: fingered 3 strands = 1 loop FD
\label{beyond}}
We present here a family, which includes almost all knots upto 10-crossings knots.
It is basically the same as the one studied in \cite{MMfam}, only we have introduced three more
parameters, which were kept fixed in that paper leading to the  10-parametric family.
This appears more efficient than the original 7-parametric one.
We refer to \cite{MMfam} for all details, and just remind the definitions.
The knot diagram is collection of seven fingers, attached to a closed 3-strand braid:

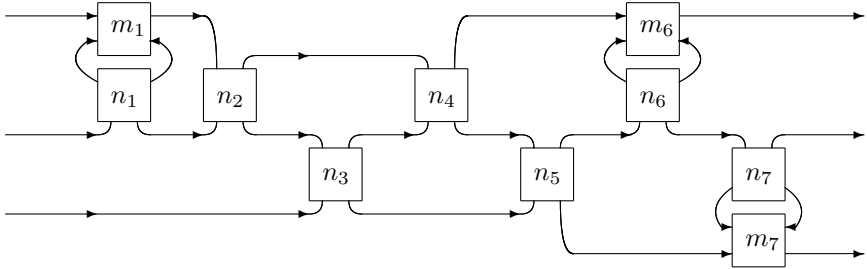
\vspace{1cm}

\begin{picture}(300,100)(-20,-50)
\put(40,45){\vector(1,0){35}}
\put(75,30){\line(0,1){20}}
\put(75,50){\line(1,0){20}}
\put(95,30){\line(0,1){20}}
\put(40,0){\vector(1,0){35}}
\put(40,-30){\vector(1,0){35}}
\put(75,-30){\vector(1,0){80}}
\qbezier(75,20)(60,28)(72,36)
\qbezier(95,20)(110,28)(97,36)
\put(73,35.5){\vector(1,0){2}}
\put(96.5,35.5){\vector(-1,0){2}}
\put(80,39){\mbox{$m_1$}}
\qbezier(75,0)(80,0)(80,5)
\put(75,25){\line(1,0){20}}
\put(75,5){\line(0,1){20}}
\put(80,12){\mbox{$n_1$}}
\put(75,5){\line(1,0){20}}
\put(95,5){\line(0,1){20}}
\put(95,45){\vector(1,0){20}}
\qbezier(90,5)(90,0)(95,0)
\put(75,30){\line(1,0){20}}
\put(95,0){\vector(1,0){20}}
\qbezier(115,45)(120,45)(120,25)
\qbezier(115,0)(120,0)(120,5)
\put(115,25){\line(1,0){20}}
\put(115,5){\line(0,1){20}}
\put(120,12){\mbox{$n_2$}}
\put(115,5){\line(1,0){20}}
\put(135,5){\line(0,1){20}}
\qbezier(130,25)(130,30)(135,30)
\qbezier(130,5)(130,0)(135,0)
\put(135,30){\vector(1,0){20}}
\put(135,0){\vector(1,0){20}}
\qbezier(155,-30)(160,-30)(160,-25)
\qbezier(155,0)(160,0)(160,-5)
\put(155,-25){\line(1,0){20}}
\put(155,-5){\line(0,-1){20}}
\put(160,-17){\mbox{$n_3$}}
\put(155,-5){\line(1,0){20}}
\put(175,-5){\line(0,-1){20}}
\qbezier(170,-25)(170,-30)(175,-30)
\qbezier(170,-5)(170,0)(175,0)
\put(155,30){\line(1,0){40}}
\put(175,-30){\vector(1,0){60}}
\put(175,0){\vector(1,0){20}}
\qbezier(195,30)(200,30)(200,25)
\qbezier(195,0)(200,0)(200,5)
\put(195,25){\line(1,0){20}}
\put(195,5){\line(0,1){20}}
\put(200,12){\mbox{$n_4$}}
\put(195,5){\line(1,0){20}}
\put(215,5){\line(0,1){20}}
\qbezier(210,25)(210,45)(215,45)
\qbezier(210,5)(210,0)(215,0)
%
\put(215,0){\vector(1,0){20}}
\qbezier(235,-30)(240,-30)(240,-25)
\qbezier(235,0)(240,0)(240,-5)
\put(235,-25){\line(1,0){20}}
\put(235,-5){\line(0,-1){20}}
\put(240,-17){\mbox{$n_5$}}
\put(235,-5){\line(1,0){20}}
\put(255,-5){\line(0,-1){20}}
\qbezier(250,-25)(250,-45)(255,-45)
\qbezier(250,-5)(250,0)(255,0)
%

\put(295,0){\vector(1,0){20}}
%


\put(275,30){\line(1,0){20}}
\put(215,45){\vector(1,0){60}}
\put(275,30){\line(0,1){20}}
\put(275,50){\line(1,0){20}}
\put(295,30){\line(0,1){20}}
\put(255,0){\vector(1,0){22}}
%
\qbezier(275,20)(260,28)(272,36)
\qbezier(295,20)(310,28)(297,36)
\put(273,35.5){\vector(1,0){2}}
\put(296.5,35.5){\vector(-1,0){2}}
\put(280,39){\mbox{$m_6$}}
\qbezier(275,0)(280,0)(280,5)
\put(275,25){\line(1,0){20}}
\put(275,5){\line(0,1){20}}
\put(280,12){\mbox{$n_6$}}
\put(275,5){\line(1,0){20}}
\put(295,5){\line(0,1){20}}
\put(295,45){\vector(1,0){70}}
\qbezier(290,5)(290,0)(295,0)

%
%

\qbezier(315,0)(320,0)(320,-5)
\put(315,-25){\line(1,0){20}}
\put(315,-5){\line(0,-1){20}}
\put(320,-17){\mbox{$n_7$}}
\put(315,-5){\line(1,0){20}}
\put(335,-5){\line(0,-1){20}}
\qbezier(330,-5)(330,0)(335,0)
%
\put(315,-30){\line(1,0){20}}
\put(315,-50){\line(1,0){20}}
\put(315,-50){\line(0,1){20}}
\put(335,-50){\line(0,1){20}}
\put(320,-42){\mbox{$m_7$}}
\put(255,-45){\vector(1,0){60}}
\put(335,-45){\vector(1,0){30}}
\qbezier(311,-35)(305,-27)(315,-20)
\put(313,-35){\vector(1,0){2}}
\qbezier(339,-35)(345,-27)(335,-20)
\put(337,-35){\vector(-1,0){2}}
\put(335,0){\vector(1,0){30}}
%
\end{picture}

\vspace{.5cm}

\noindent
Here $m_{1,6,7}$, $n_1$ and $n_6$ are even, the other five parameters  $n_{2,3,4,5}$ and $n_7$ are odd.
In the 7-parametric family of \cite{MMfam} the three $m$'s were fixed to be $2$,$2$,$\pm 2$. The fingers are
\be
P^{(n_{2,3,4,5})}_X =  \frac{(\bar S\bar T^{n_{2,3,4,5}}S)_{_{\vac,X}}}{S_{_{\vac, X}}}\nn\\
K^{(m_{1,6},n_{1,6})}_X = \frac{( S T^{m_{1,6}} S^\dagger \bar T^{n_{1,6}}S)_{_{\vac, X}}}{S_{_{\vac, X}}}\nn\\
\bar K^{(m_7,n_7)}_X = \frac{(\bar S\bar T^{m_7}\bar S \bar T^{n_7}S)_{_{\vac, X}}}{S_{_{\vac, X}}}
\label{7parfam}
\ee
This picture of the knot is rather symbolic, since one has also to mark the way how the small loops nearby
the boxes $n_1$, $n_6$ and $n_7$ cross the strands. It can be read off from the formula that is really used for the calculation:
in the case of the fundamental representation $R=[1]$:
$$
d_{[1]}H_{[1]}^{(n_1,\ldots,n_7|\pm)} \ \ = \ \ d_{[3]} \cdot
K^{(m_1,n_1)}_{[2]}\cdot \left(
\prod_{i=2}^{5}   P^{(n_i)}_{[2]}\right) K^{(m_6,n_6)}_{[2]}\bar K^{(n_6,n_7)}_{[2]}
\ \ + \ \ d_{[111]}\cdot    K^{(m_1,n_1)}_{[11]}\cdot \left(
\prod_{i=2}^{5}   P^{(n_i)}_{[11]}\right) K^{(m_6,n_6)}_{[11]}\bar K^{(m_7,n_7)}_{[11]} \ \ +
$$
\vspace{-0.3cm}
\begin{multline}
+\ \ d_{[21]}\cdot  \Tr_{2\times 2} \left\{
\left(\begin{array}{cc} K^{(m_1,n_1)}_{[2]} & 0 \\ \\  0 & K^{(m_1,n_1)}_{[11]} \end{array}\right)
\left(\begin{array}{cc} P^{(n_2)}_{[2]} & 0 \\ \\  0 & P^{(n_2)}_{[11]} \end{array}\right)
\left(\begin{array}{cc} \frac{1}{[2]}& \frac{\sqrt{[3]}}{[2]} \\ \\
\frac{\sqrt{[3]}}{[2]} & -\frac{1}{[2]} \end{array}\right)
\left(\begin{array}{cc} P^{(n_3)}_{[2]} & 0 \\ \\  0 & P^{(n_3)}_{[11]} \end{array}\right)
\left(\begin{array}{cc} \frac{1}{[2]}& \frac{\sqrt{[3]}}{[2]} \\ \\
\frac{\sqrt{[3]}}{[2]} & -\frac{1}{[2]} \end{array}\right)  \cdot
\right.
\\
 \cdot\left(\begin{array}{cc} P^{(n_4)}_{[2]} & 0 \\ \\  0 & P^{(n_4)}_{[11]} \end{array}\right)
\left(\begin{array}{cc} \frac{1}{[2]}& \frac{\sqrt{[3]}}{[2]} \\ \\
\frac{\sqrt{[3]}}{[2]} & -\frac{1}{[2]} \end{array}\right)
\left(\begin{array}{cc} P^{(n_5)}_{[2]} & 0 \\ \\  0 & P^{(n_5)}_{[11]} \end{array}\right)
\left(\begin{array}{cc} \frac{1}{[2]}& \frac{\sqrt{[3]}}{[2]} \\ \\
\frac{\sqrt{[3]}}{[2]} & -\frac{1}{[2]} \end{array}\right)
\left(\begin{array}{cc} K^{(m_6,n_6)}_{[2]} & 0 \\ \\  0 & K^{(m_6,n_6)}_{[11]} \end{array}\right)
\cdot
\\
\cdot
\left.
\left(\begin{array}{cc} \frac{1}{[2]}& \frac{\sqrt{[3]}}{[2]} \\ \\
\frac{\sqrt{[3]}}{[2]} & -\frac{1}{[2]} \end{array}\right)
\left(\begin{array}{cc} \bar K^{(m_7,n_7)}_{[2]} & 0 \\ \\  0 & \bar K^{(m_7,n_7)}_{[11]} \end{array}\right)
\left(\begin{array}{cc} \frac{1}{[2]}& \frac{\sqrt{[3]}}{[2]} \\ \\
\frac{\sqrt{[3]}}{[2]} & -\frac{1}{[2]} \end{array}\right)
 \right\} \ \ \ \ \ \ \ \
\end{multline}

\bigskip

\noindent
This 10-parametric family is rather rich --
from Rolfsen table it misses at most {\it three} arborescent knots
$10_{79}$, $10_{99}$, $10_{152}$ and {\it twelve} non-arborescent knots:
$8_{18}$, $10_{109},10_{112},10_{114}-10_{116},10_{118}, 10_{120}-10_{123},10_{163}$.
For concrete values of parameters, associated with particular knots see \cite{knotebook}.

\section*{Acknowledgements}

\noindent

We are indebted to Andrei Malutin for teaching us a lot about arborescent knots. We are also grateful to Saswati Dhara for checking some examples.

Our work is partly supported by the Indian-Russian grant: 14-01-92691-Ind-a; INT/RFBR/P-162. It is also partly supported
by RFBR grants 16-01-00291 (A.Mir.), 16-02-01021 (A.Mor.), mol-a-dk 16-32-60047 (And.Mor) and by grant ÌÊ-8769.2016.1 (A.S.). Also we are partly supported by the Quantum Topology Lab of Chelyabinsk State University (Russian Federation government grant 14.Z50.31.0020) (An.Mor. \& A.S.).

\end{document}